\documentclass[twocolumn,aps,prl,10pt,amsmath,amssymb,nofootinbib,showpacs,superscriptaddress,floatfix]{revtex4-1}

\DeclareFontFamily{U}{rcjhbltx}{}
\DeclareFontShape{U}{rcjhbltx}{m}{n}{<->rcjhbltx}{}
\DeclareSymbolFont{hebrewletters}{U}{rcjhbltx}{m}{n}

\usepackage{graphicx}
\usepackage{color}
\usepackage[usenames,dvipsnames]{xcolor}
\usepackage[colorlinks=true,linkcolor=Red,citecolor=Green,linktoc=page]{hyperref}
\usepackage{multirow}
\usepackage{float}
\usepackage{flushend}
\usepackage{balance}
\usepackage[varg]{txfonts}
\usepackage{ulem}
\usepackage{fancyhdr}

\DeclareMathSymbol{\lamed}{\mathord}{hebrewletters}{108}

\begin{document}
\title{%Dynamic critical exponents for the current response of superinsulators
Relaxation electrodynamics of superinsulators}

	\author{A.\,Mironov}
	\affiliation{Terra Quantum AG, St. Gallerstrasse 16A, CH-9400 Rorschach, Switzerland}

	\author{M.\,C.\,Diamantini}
	\affiliation{NiPS Laboratory, INFN and Dipartimento di Fisica e Geologia, University of Perugia, via A. Pascoli, I-06100 Perugia, Italy}

\author{C.\,A.\,Trugenberger}
\affiliation{SwissScientific Technologies SA, rue du Rhone 59, CH-1204 Geneva, Switzerland}

\author{V. M. Vinokur\,$^1$}

\vspace{-0.9cm}
\date{\today}
\begin{abstract}
Superinsulators offer a unique laboratory realizing strong interaction phenomena like confinement and asymptotic freedom in quantum materials. Recent experiments
	evidenced that superinsulators are the mirror-twins of superconductors with reversed electric and magnetic field effects. Cooper pairs and Cooper holes in the superinsulator are confined into neutral  electric pions by electric strings, with the Cooper pairs playing the role of quarks. Here we report the non-equilibrium relaxation of the electric pions in superinsulating films. We find that the time delay $t_{\mathrm{sh}}$ of the current passage in the superinsulator is related to the applied voltage $V$ via the power law,  $t_{\mathrm{sh}}\propto (V-V_{\mathrm p})^{-\mu}$, where  $V_{\mathrm p}$ is the effective threshold voltage. Two distinct critical exponents, $\mu=1/2$ and $\mu=3/4$, correspond to jumps from the electric Meissner state to the mixed state and to the superinsulating resistive state with broken charge confinement, respectively. The $\mu=1/2$ value establishes a direct experimental evidence for the electric strings' linear potential confining the charges of opposite signs in the electric Meissner state and effectively rules out disorder-induced localization as a mechanism for superinsulation.
	We further report the memory effects and their corresponding dynamic critical exponents arising upon the sudden reversal of the applied voltage. Our observations open routes for exploring fundamental strong interaction charge confinement via desktop experiments.

\end{abstract}
\maketitle

\vspace{-0.9cm}

%\begin{bibunit}[plain]
%\section{Introduction}
\noindent

\section*{Introduction}
\noindent
Superinsulators are materials dual to superconductors, i.e., materials that at low but finite temperatures have infinite resistance so that no electric current can passes through them. Superinsulation arises in materials in which an effective electromagnetic action contains magnetic monopoles, so that Coulomb interactions dominate magnetic interactions 
and offers a unique laboratory realizing strong interaction phenomena like confinement and asymptotic freedom in quantum materials. Recent experiments\,\cite{electrostatics}
revealed that superinsulators reverse electric and magnetic field effects with respect to superconductors so that Cooper pairs and Cooper holes in the superinsulator are confined into neutral  electric pions by electric strings, with the Cooper pairs playing the role of quarks\,\cite{electrostatics,dtv1}.
The superconductor-to-insulator transition (SIT) is a paradigmatic quantum phase transition \cite{efetov} driven by the strength of the Coulomb interaction in thin superconducting films, which can be tuned by the film thickness\,\cite{haviland}, magnetic field \cite{fisher1, fisher2, fisher3}, gate voltage\,\cite{marcus}, and disorder\,\cite{finkelstein}. The competition between magnetic and electric forces gives rise to the superconductor-superinsulator\,\cite{dst, vinokurnature, vinokurannals} transition that may occur via an intermediate Bose metal phase\,\cite{dst, das1, das2, phillips1, phillips2, kapitulnik}, shown to be a bosonic topological insulator (BTI)\,\cite{bm}. At low temperatures, the BTI behaves as a metal, with currents carried by the topologically protected edge excitations. At temperatures comparable to the topological Chern-Simons gap\,\cite{jackiw1, jackiw2}, the BTI displays activated, insulating behaviour. It is this behaviour that was observed in the first experiments \cite{haviland} and coined the term SIT. 

The superinsulating state is a confinement phase\,\cite{polyakov} where Cooper pairs and Cooper holes are bound by electric strings providing a linear attractive potential between them and where Coulomb interactions are screened on scales much smaller than the typical string length\,\cite{dtv1}. The superinsulator can be destroyed either by a sufficient temperature increase \cite{vft} or by applying a sufficiently high voltage. The latter is the effect dual to destroying superconductivity by applying a strong magnetic field\,\cite{tinkham}. Mirroring type II superconductors via the electric and magnetic field effects reversal, there are two critical voltages, $V_{\rm c1}$ and $V_{\rm c2}$ (`substituting' the lower, $H_\mathrm{c1}$ and upper, $H_\mathrm{c2}$, magnetic critical fields). At $V<V_{\rm c1}$ the superinsulator is in its electric Meissner state where applied electric fields are expelled\,\cite{electrostatics}. In this state the current response is exponentially suppressed and vanishes in the infinite-sample limit, giving rise to the infinite resistance at finite temperatures. At $V_{\rm c1} < V < V_{\rm c2}$, the electric field penetrates the sample in a form of thin electric flux filaments, Polyakov strings, and the current starts passing through. In this state, the $I(V)$ characteristics show a power law\,\cite{electrostatics}. Finally, voltages $V\geqslant V_{\rm c2}$ destroy the superinsulating state, and the superinsulator transforms into the resistive bosonic topological insulator (BTI)\,\cite{electrostatics,dtv1}. 

The two critical voltages appear as two kinks in the $I(V)$ characteristics\,\cite{electrostatics}. However in the low-resistive samples near the SIT,  $V_{\rm c1}$ and $V_{\rm c2}$ may often appear very near to each other, and experiments performed in this region can fail to separate them. The corresponding observation of only one kink has led to a speculation\,\cite{overheating, shahar} that the superinsulating behavior  might arise due to disorder-induced-localization\,\cite{mbl}. This idea originated from the partial explanation of the shape of the jump in the $I(V)$ curves by charge overheating \cite{overheating}, which was erroneously taken as a model of the nature of the superinsulating state, while it has only to do with one possible mechanism by which this state is dynamically destroyed. Here we report the first ever experiment on the dynamical behaviour of superinsulators and we show that, even when the two critical voltages are so close that they cannot be distinguished in the $I(V)$ curves, they are clearly identified by their different dynamic critical exponents relating the delay in the current response to a sudden voltage pulse, $t_{\mathrm{sh}}\propto (V-V_{\mathrm p})^{-\mu}$, where  $V_{\mathrm p}$ is the effective threshold voltage. Furthermore, we show that the exponent $\mu=1/2$ describing the jump from the Meissner state to the mixed state is a direct confirmation of the linear binding potential between charges caused by electric strings in the confined state, while the second exponent $\mu=3/4$ reflects the quantum phase slips in the 1D edge channel of the intermediate bosonic topological insulator phase separating superinsulation from superconductivity. These results fully support the monopole-induced confinement nature of superinsulators and strongly disfavour the assumption that disorder-induced-localization may model the superinsulating state.

%%%%%%%%%%%%%%%%%%%%%%%%%%%%%%%%%%%%%%%%%%%%
\begin{figure*}
	\begin{center}
		\includegraphics[width=0.8\linewidth]{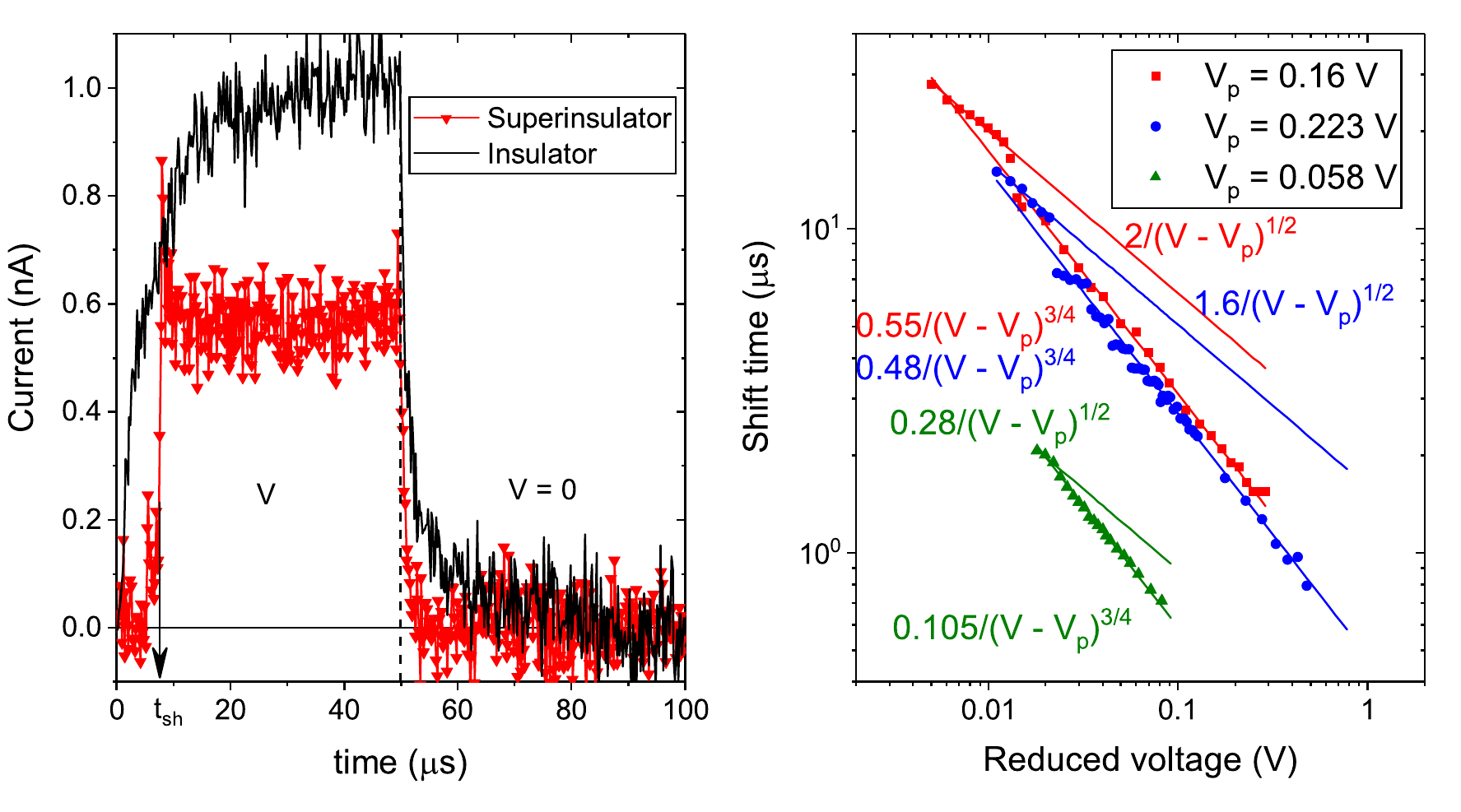}	
		\caption{\textbf{Relaxation response.} \textbf{a}. Characteristic dependence of current on time when a rectangular voltage pulse with amplitude $V=0.155$\,V and duration 50 $\mu$s is applied at temperature $T=$20\,mK when measuring the superinsulator and $T=$300\,mK, when measuring the insulator. \textbf{b}. Scaling of rising edge shift time $t_\mathrm{sh}$ dependence
			upon amplitude of rectangular voltage pulse (symbols). The solid lines show the fitting functions.} \label{fig_exp}
	\end{center}
\end{figure*}

\begin{figure}
	\begin{center}
		\includegraphics[width=0.8\linewidth]{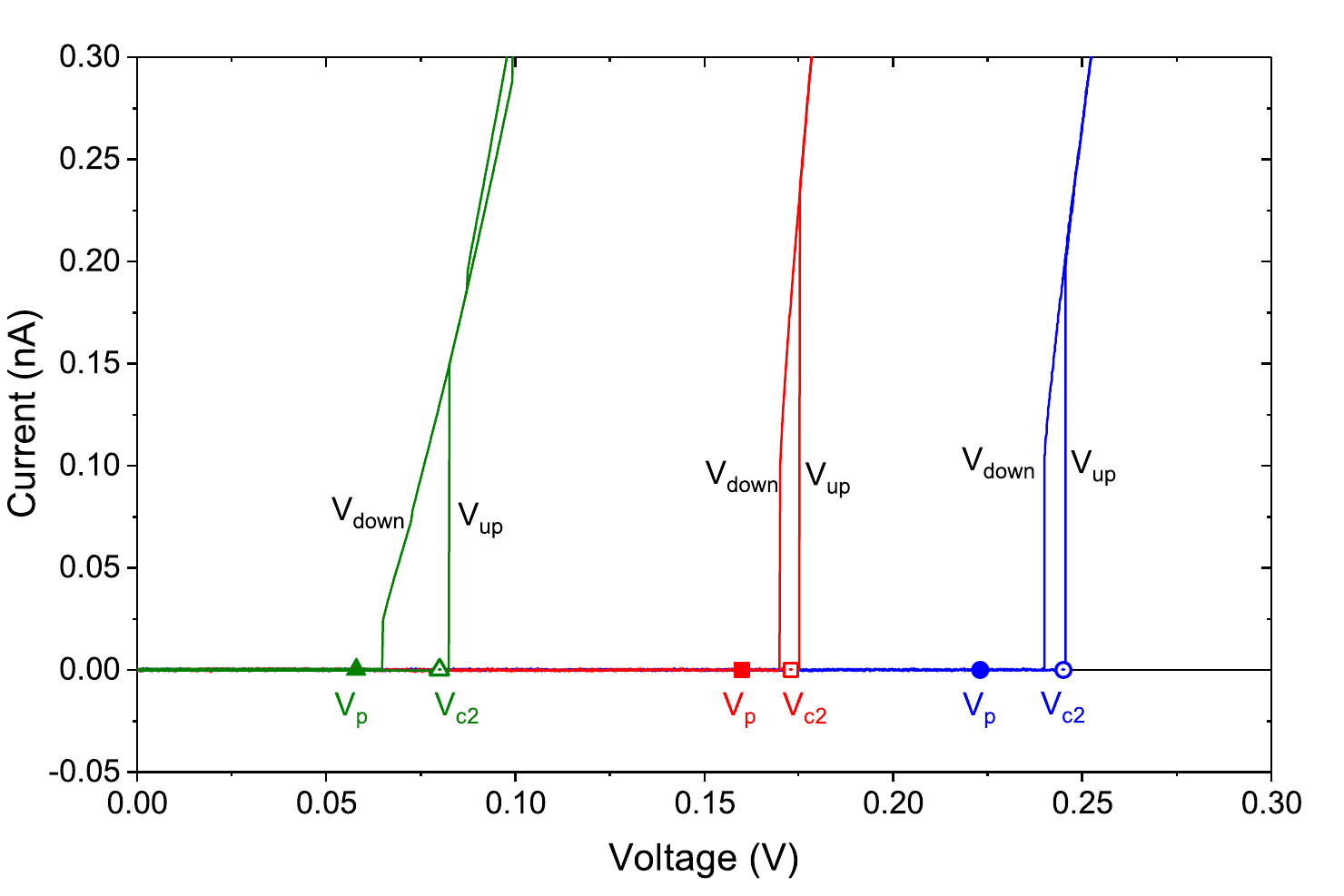}	
		\caption{\textbf{Current-voltage characteristics.} Different color correspond to different samples. Each sample demonstrates the hysteretic behavior around the threshold voltage $V_{c2}$, the black arrows indicate the direct and reversed directions of the voltage change. The filled symbols show the position of $V_{\rm p}$ relative to the threshold voltage $V_{c2}$, presented by empty symbols.} \label{fig_VAch}
	\end{center}
\end{figure}

%%%%%%%%%%%%%%%%%%%%%%%%%%%%%%%%%%%

\section*{Experiment}
The measurements are taken on the superinsulating NbTiN films grown by the atomic layer deposition (ALD) technique. The low-temperature experiments are carried out in a $^3$He/$^4$He dilution refrigerator Triton 400 equipped with the measuring RF lines. To measure the current when the low frequency, $2 \times 10^{4}$ Hz, pulsed voltage is applied, the measuring circuit consisting of an Agilent 81150A pulse generator, a Femto HCA high-speed current amplifier with a bandwidth of 0 - 2 MHz, and a Tektronix DPO 7354C oscilloscope, is used. The time of the voltage rise/fall in this circuit does not exceed 500\,ns. %The low- and high-pass filters are not used for minimization influence of measurement circuit on measurement current.

\section*{Results}
Figure \ref{fig_exp}a displays a typical response of the current in a superinsulating film to a square pulse voltage, as compared to the dynamic response of the normal insulating state at higher temperatures. Contrary to the normal insulator, in which the current increases continuously to the steady state, in the superinsulator at the first moment, no measurable current value appears in the system. Then, after some delay time, $t_\mathrm{sh}$, a measurable current arises and quickly reaches saturation. When the voltage is turned off, the current decreases exponentially. These features of the current response are observed provided the pulse duration is at least 1.5 times longer than the time $t_\mathrm{sh}$.
In Figure\,\ref{fig_exp}b we present the dependencies of the shift time of the rising edge $t_\mathrm{sh}$ upon the amplitude $V$ of the pulse signal for several samples which differ from each other in resistivity and hence in the resulting threshold voltages. Note that with the increase in resistivity, the increase in the threshold voltage is observed.
As the threshold voltage grows, the rising edge shift time first increases, then reaches a maximum, and then, finally, begins to decrease. We find that the shift time $t_\mathrm{sh}$ is a power law function of the reduced voltage $V-V_\mathrm{p}$,  i.e., $t_\mathrm{sh}\propto(V-V_\mathrm{p})^{-\mu}$, where $V_\mathrm{p}$ is a fitting parameter which depends only upon the sample. At some voltage, the power $\mu$ abruptly changes. In addition, at voltages less than $V_\mathrm{c2}$, the effect of shifting the front edge of the pulse and measurable current level are observed only when the signal is periodic with the period being less than\,1\,ms.
The small AC component promotes the depinning of the charges at the ends of electric flux tubes in an exact analogy to the AC current component promoting the vortex mobility and the resistive behaviour in the mixed state of superconductors in the experiments on superconductors\,\cite{tinkham}.
The voltage $V_{\mathrm c2}$ corresponds to the second threshold voltage of the film\,\cite{electrostatics}, see Fig.\,\ref{fig_VAch}. At the same time, the voltage $V_{\mathrm p}$ is smaller than the threshold voltage, $V_{\mathrm c2}$, which has a hysteretic behavior presented in Fig.\,\ref{fig_VAch}. As we will show in the next section, $V_{\mathrm p}$ must be identified with the first threshold voltage $V_{\mathrm c1}$.

\begin{figure*}
	\begin{center}
		\includegraphics[width=0.8\linewidth]{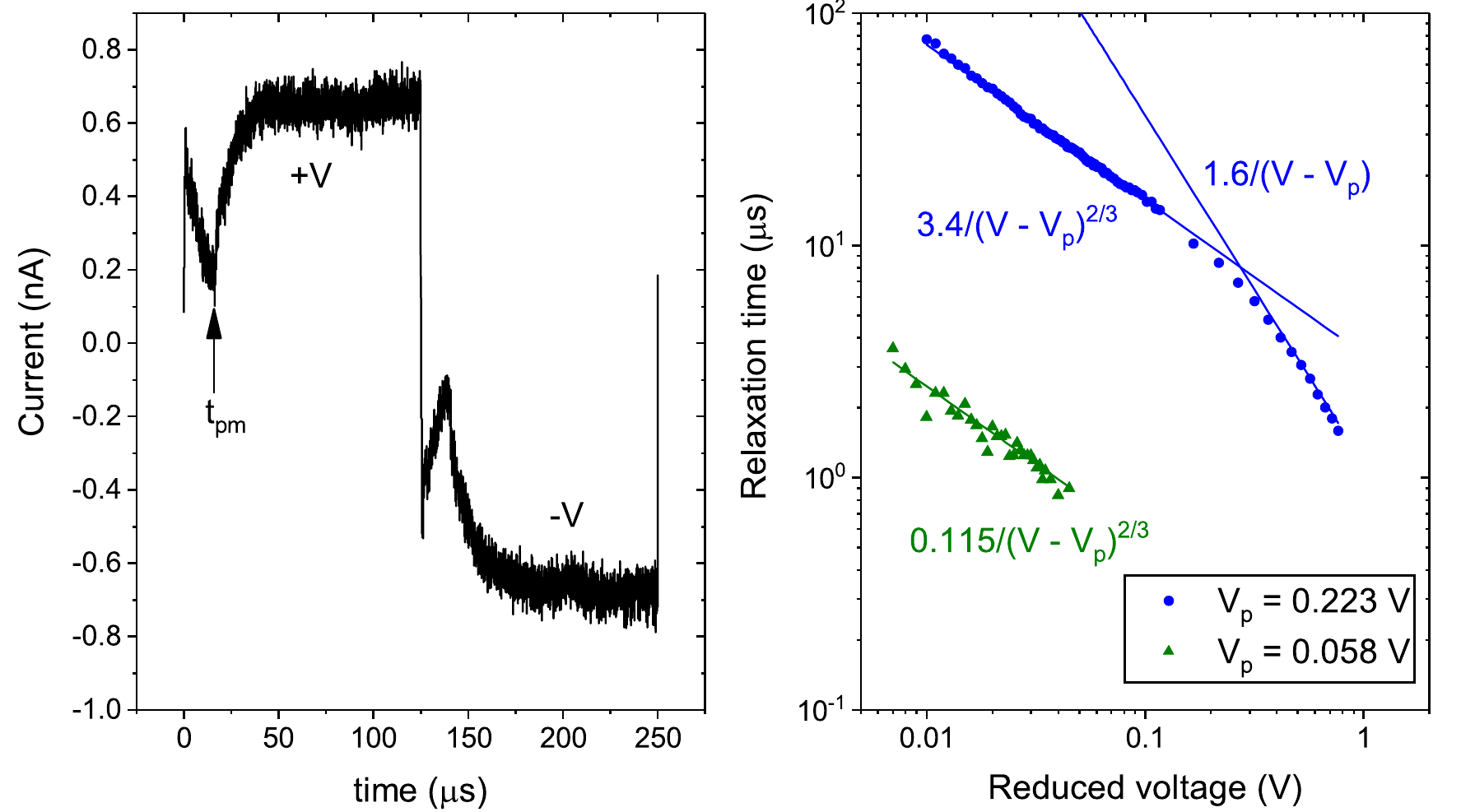}	
		\caption{\textbf{Relaxation response upon reversing voltage.} \textbf{a}. Characteristic dependence of current on time when a rectangular voltage pulse with amplitude $\pm V_{\rm m}$ and period 125 $\mu$s is applied at a temperature of 20 mK. \textbf{b}. Scaling of local minima time $t_\mathrm{pm}$ dependence
			from amplitude of rectangular voltage pulse (symbols). The solid line shows the fitting functions.} \label{fig_pm_exp}
	\end{center}
\end{figure*}

Figure\,\ref{fig_pm_exp} shows similar results for voltage jumps from $+V$ to $-V$. Unlike the above $0\to |V|$ jumps, now as the sign of the voltage changes, the current also instantly changes its sign. However, in the further time interval, $t_\mathrm{pm}$, the current decreases and, having reached its minimum depending on the voltage value, again rapidly increases to saturation. Thus, when the sign of the voltage changes, the ``initial" dynamic state is first destroyed and only then a new state is formed. The time $t_\mathrm{pm}$ is also a power function of the reduced voltage $V-V_\mathrm{p}$, with the same probe voltage but with another exponent. Note that, when the sign of the voltage changes, there are no features associated with the threshold voltage $V_\mathrm{c2}$. At the same time, at voltages $V$ between $V_\mathrm{p}$  and $V_\mathrm{c2}$ the requirement for a signal to be periodic with a period of less than 1 ms is persisting.

%%%%%%%%%%%%%%%%%%%%%%%%%%%%%%%%%%%
%%%%%%%%%%%%%%%%%%%%%%%%%%%%%%%%%%%%

\section*{Linear potential and confinement by electric strings}
Now we demonstrate how the observed relaxation behavior results from the linear string confinement. Consider two charges $\pm 2e$ of mass $m$ subject to the screened Coulomb interactions and bound by the linear potential $V\left( {\bf x}_1, {\bf x}_2 \right) =\sigma |{\bf x}_1-{\bf x}_2|$, where $\sigma$ is the string tension of the confining electric string and ${\bf x}_i$ are the coordinates of the two particles. The critical value of the potential when current starts to pass is $V_{\rm c1}= \sigma L$, with $L$ being the sample length, corresponding to a configuration in which the two opposite charges reach the sample boundaries. 

In center of mass coordinates we can reduce the problem to one particle of the reduced mass $m/2$ at the origin, subject to the potential energy $U(r) = 2e \sigma r$, with $r$ denoting the separation distance. The binding energy of the $\pm$ charge pair is 
$2e V_{\rm c1} \approx 0.3$ eV, where we used the average value of the threshold voltages in Fig. 1. This is far smaller than the reduced mass 0.5 MeV of a Cooper pair. We can thus safely use classical considerations to compute the behaviour of the centre of mass coordinate. A positive linear potential results in an attractive constant force $F_{\mathrm a}= 2e \sigma = 2e V_{\rm c1}/L$. Let us now apply an external potential $V$. This amounts to an additional repulsive force $F_{\rm r} = 2e V/L$. The total acceleration of separation is thus $a=(2/m) F_{\rm tot} = (4e/mL) (V-V_{\rm c1})$. The corresponding equation of motion separating the charge-hole system is
\begin{equation}
	r(t) = \frac{2e}{ mL} \left( V-V_{\rm c1} \right) t^2 \ .
	\label{eqmotion}
\end{equation}
Current starts to pass when the charges of opposite sign reach the sample boundaries, i.e. when $r\left( t_{\rm cr} \right) = L$. The delay for current passage is thus given by inverting Eq.\,(\ref{eqmotion}),
\begin{equation}
	t_{\rm cr} = \sqrt{mL^2\over 2e} \left( V- V_{\rm c1} \right)^{-1/2} \ .
	\label{crittime}
\end{equation}
This shows that the observed value $\mu=1/2$ of the dynamical critical exponent  for the current passage is a direct consequence of the linear binding potential between charges of the opposite sign and that $V_{\mathrm p}$ must be identified with the first critical voltage $V_{c1}$, thereby fully confirming the confinement nature of superinsulators. Note, that even if the $I(V)$ measurements are not accurate enough to disentangle the two critical voltages, they can still be clearly identified by their two different dynamic critical exponents.

\section*{Quantum dissipation}
For higher applied voltages the system jumps directly over the second critical voltage $V_{\rm c2}$ beyond which the superinsulation is completely destroyed. Had the resulting state at temperature $T=20$\,mK, been an Ohmic resistive state, where it is the velocity of charges which is proportional to the applied force, rather than the charges' acceleration, the above simple calculation shows that we would have expected the critical exponent $\mu=1$, which is clearly not the case. Instead, for NbTiN, this state is a BTI\,\cite{bti1, bti2, bti3}, with a topological gap for bulk excitations and charge transport mediated by the symmetry-protected edge modes\,\cite{bm,ournodisorder}. Only at higher temperatures one would have observed a thermally activated bulk conductance. At the lowest temperatures, we expect that the quantum dissipation occurs only due to quantum phase slips\,\cite{zaikin, choi, larkin} across the 1D edge conducting channels.

The temperature-independent but velocity-dependent friction is, indeed, a fundamental characteristic of the 1D bosonic conduction channels in the quantum phase slips regime\,\cite{danshita}. Such a friction has recently been experimentally confirmed, although there is not yet full quantitative agreement with theory\,\cite{tanzi}. The velocity dependent damping coefficient implies an anomalous relation
\begin{equation}
	F \propto v^{2K-2} \propto \Gamma(v)
	\label{anom}
\end{equation}
between the applied force $F \propto\left( V- V_{\rm cr}\right) $ and the velocity, with $K$ being the Luttinger parameter of the bosons, see\,\cite{luttinger} for a review. This is the same power law that has been originally obtained in\,\cite{golubev}, where it was shown that the quantum phase slip nucleation rate $\Gamma$ plays the role of the applied force in the quantum wire dissipation. Repeating the above calculation we obtain an arrival time scaling
\begin{equation}
	t_{\rm cr} \propto \left( V- V_{\rm cr} \right)^{-{1\over 2K-2} } \ ,
	\label{anarr}
\end{equation}
which matches the experimental result for $K=5/3$.

\section*{Switching delay} 
Let us now address the delay $t_{\pm}$ observed upon the sudden switching the direction of the applied electric field. Suppose that we start off in a situation with the negative voltage having the magnitude larger than $V_{c2}$ and therefore, the corresponding steady-state current goes in the negative direction. When switching to the positive voltage of the same magnitude the system will momentarily pass through the mixed and Meissner states of the superinsulator. In this process, the strings form, which are initially taut in the negative direction. But when the applied field passes through the zero, the charges can flow freely in the opposite direction for a while, until the string completely ``turns around", i.e., until the tension builds up in the opposite direction. This explains the initial bump in the current, which then decreases when the string tension increases to oppose the inverted applied force. Since the applied voltage is larger than $V_{\mathrm c2}$, at a certain point the mixed state is again destroyed to form a steady current, now flowing in the positive direction. The delay $t_{\mathrm{pm}}$ is thus another, albeit qualitative, confirmation of the string-confinement picture of the superinsulation. A quantitative derivation of the observed scaling of $t_{\mathrm{pm}}$ with the power $\mu=2/3$ is beyond the present theoretical framework.

\section*{Discussion and conclusion}
\noindent
The results of our measurements of dynamical exponents describing the relaxation of the current passage confirm the existence of two electric states of superinsulators, the Meissner state, where the electric field is expelled from the sample and the mixed state where electric field penetrates superinsulator in  form of the electric filaments or electric strings. These two states of a superinsulator constitute an exact duality to the conventional Meissner and mixed states in a superconductor. Furthermore, the exponent $\mu=1/2$ for the transition from the Meissner to the mixed state provides a direct validation of the linear binding potential created by the electric strings between charges of opposite sign in the electric Meissner state. Our results rule out any superinsulation model not containing  this linear binding potential preventing charge transport at voltages not exceeding the critical voltage. In particular, it strongly disfavors disorder-induced localization as a mechanism for current suppression and hyperactivation.

\subsubsection*{Data availability}
\noindent
The data that support the findings of this study are available from the corresponding authors upon reasonable request.
%Example text under a subsection. Bulleted lists may be used where appropriate, e.g.

%\begin{itemize}
%\item First item
%\item Second item
%\end{itemize}

%\subsubsection*{Third-level section}
 
%Topical subheadings are allowed.

%\section*{Discussion}
%The Discussion should be succinct and must not contain subheadings.

%\section*{Methods}

%\bibliography{sample}

%\noindent LaTeX formats citations and references automatically using the bibliography records in your .bib file, which you can edit via the project menu. Use the cite command for an inline citation, e.g.  \cite{Hao:gidmaps:2014}.

%For data citations of datasets uploaded to e.g. \emph{figshare}, please use the \verb|howpublished| option in the bib entry to specify the platform and the link, as in the \verb|Hao:gidmaps:2014| example in the sample bibliography file.

\section*{Acknowledgements}
	The work of A.M. and V.M.V. was supported by Terra Quantum AG.

\section*{Author contributions statement}
	A.M., M.C.D.,  C.A.T., and V.M.V conceived the work, A.M. carried out the experiment, M.C.D.,  C.A.T., and V.M.V carried out the calculations, all authors discussed the results and wrote the manuscript.

\section*{Additional information}
%To include, in this order: \textbf{Accession codes} (where applicable); 
\textbf{Competing interests} The authors declare no competing interests. \\

\end{document}